# Universal method of selective detection of a wide range of pollutants in liquids using conductance quantization

O. Pospelov[1], A. Herus[2], A. Savytskyi[2], V. Vakula[2],
M. Sakhnenko[1], N. Kalashnyk[3], E. Faulques[3], G. Kamarchuk[2*]

1 – National Technical University "Kharkiv Polytechnic Institute", 2 Kyrpychov Str., Kharkiv, 61002, Ukraine.
2 – B. Verkin Institute for Low Temperature Physics and Engineering, 47 Nauky Ave., Kharkiv, 61103, Ukraine.
3 – Univ. Lille, CNRS, Centrale Lille, Yncréa ISEN, Univ. Polytechnique Hauts-de-France, UMR 8520-IEMN, F-59000, Lille, France

*email: kamarchuk.iltpe@gmail.com

## Abstract

The primary objective of research in modern sensor technologies is to develop innovative detection methods for the rapid analysis of complex molecular systems. The present work demonstrates that the quantum mechanism of selective detection, based on conductance quantization, can be effectively employed to create a universal method for detecting a broad spectrum of agents in liquid media, including heavy metals and organic solvents. The efficacy of this approach is illustrated through the use of quantum point-contact sensors, which utilize dendritic Yanson point contacts undergoing quantum transformations during the cyclic switchover effect. These sensors have proven capable of detecting copper, zinc, and lead ions in liquid media across a wide range of concentrations, including trace levels as low as a few parts per billion (ppb). Furthermore, they can identify organic solvents, as demonstrated with acetic acid. The use of innovative quantum detection principles paves the way for the development of a comprehensive array of next-generation devices, offering promising solutions for advanced environmental monitoring applications.

## 1. Introduction

The primary objective of research in modern sensor technologies is to develop novel detection methods for rapid analysis of complex molecular systems. To achieve this goal, a wide variety of sensor devices have been developed, based on various principles of operation [1]. In most cases, these devices operate by detecting electrical signals that arise from changes in the composition or structure of a system, where a sensing element interacts with the analysed environment. Sensors that provide a response to external influences in the form of a change in the electrical resistance of the sensing element have become widely used [2-7]. The response function of such transducers typically exhibits a curve with a local extremum, the magnitude of which correlates with the value of the parameter being measured. Sensors that analyse this extremum are primarily used for specific tasks, such as detecting individual substances. A more advanced device of this type is the so-called electronic nose, which consists of multiple sensing elements, each selectively responsive to a particular component in the analysed environment. This configuration enables the analysis of complex, multicomponent mixtures [8], including real-world environmental objects [9, 10].

The use of several parallel-connected autonomous sensing elements in electronic nose devices, which collectively provide a more comprehensive and detailed characterization of the environment, represents an extensive approach to advancing sensor technology: the greater the number of elements, the more information is gathered. However, alongside this extensive approach, a method based on intensive principles is also employed in sensor analysis. This method involves the active (multifaceted) interaction of a single sensing element with the studied object. By varying the modes and profiles of the signal applied to the object (essentially conducting a multifaceted inquiry) and analysing the resulting responses, one can thoroughly investigate the behaviour of the system under different conditions [11, 12]. This intensive approach is especially prevalent in modern medical diagnostics. For instance, alternating current [13] and pulse [14] stimuli, followed by systematic processing of the electrical responses, enable highly reliable non-invasive monitoring of various organs and regions of the human body.

An essential aspect of advancing new sensor technologies is the identification and development of materials with novel sensing properties. Research in this area is progressing in two primary directions: synthesizing metal alloys, creating and testing nanostructured semiconductor oxides [15, 16], and synthesizing low-dimensional nanomaterials with unique sensing characteristics, such as polymers, aptamers, graphene, nitrides, and transition metal carbides (MXenes) [17-22]. Developing new sensing materials is a complex and time-consuming task that requires significant resources, including expensive equipment and intricate technological processes.

Currently, one of the most pressing challenges in sensor science and technology is the analysis of liquid media [5, 23-26]. This has spurred numerous studies aimed at understanding the sensing phenomena in liquid environments and developing innovative detection methods [27-30]. The outcome of these studies is a broad range of techniques for analysing liquid media, utilizing scientific and technical solutions based on various physical effects and cutting-edge technologies. These include a variety of optical spectroscopic methods, such as vibrational spectroscopy, ultraviolet-visible (UV-Vis) spectrometry, absorption spectrometry, fluorescence spectroscopy, surface-enhanced Raman spectroscopy, and others [31, 32]. Mass spectrometry and liquid chromatography provide high sensitivity in detecting pollutants in liquid media [33, 34]. However, the main drawbacks of these techniques are their high cost, equipment complexity, and the need for highly trained personnel, driving the search for alternative solutions that offer greater economic efficiency, simplicity, real-time analytical depth, and portability. As a result, modern sensor technologies are emerging as strong contenders in this field. Among these, electrochemical analysis methods hold significant potential due to their advantages in cost-effectiveness and practicality [35-37].

A common characteristic of traditional sensors, which rely on the electrical conductance of sensing elements as their information signal, is their operation based on classical physics principles. While effective, this limits their range of capabilities and applications. New opportunities for developing innovative detection methods based on changes in electrical conductance arose following the discovery of quantum mechanisms for detection in complex media [38]. Quantum detection mechanisms can exploit the energetic properties of objects for recognition. Each quantum system is defined by a specific set of metastable states with distinct energy values. These metastable states, along with their associated energy values, serve as markers that form a unique "energetic fingerprint" for the system, analogous to how a human fingerprint uniquely identifies an individual. This energetic principle provides a groundbreaking solution, enabling the successful development and implementation of advanced sensor devices that operate according to quantum mechanical laws and offer universal selective detection capabilities. The practical application of this energetic approach has been demonstrated through quantum point-contact sensors [39].

A distinctive characteristic of quantum detection mechanisms is the high information content in the response from quantum sensors, enabling the development of numerous independent methods for selective detection based on a single output characteristic, such as the temporal dependence of the electrical conductance of the sensing element. This capability has been effectively demonstrated in the detection and analysis of human breath, one of the most complex biological gas media known. The energetic profile of human breath, recorded with a quantum point-contact sensor, can be decoded in real time using several alternative methods, including the integrated determination of interaction energies between the sensor and the target gas component [40], the integrated analysis of response signal markers [41], or the identification of a specific response signal marker [42]. These approaches significantly enhance the reliability of medical diagnostics related to human health conditions [39].

The implementation of quantum detection mechanisms can be further advanced by incorporating novel quantum effects [40, 43]. One such effect is the cyclic switchover effect [44], which is associated with quantum transformations in the structure of dendritic point-contact nanosystems under an electric field. This effect manifests as quantum processes of electrochemical growth and dissolution of metallic dendrites, accompanied by the synthesis and destruction of Yanson point contacts [39, 45]. Yanson point contacts are unique nanostructures with quantum properties that facilitate the development of a range of innovative technologies, including Yanson point-contact spectroscopy [45, 46] and quantum point-contact sensorics [47, 48]. The cyclic switchover effect is useful for demonstrating the quantum properties of these nanoobjects through their electrical conductance and can be an efficient tool for registering various quantum states of the point-contact system in dynamic mode [43]. This effect has considerable potential for creating universal quantum methods for selective detection in complex environments. Previous research has already shown the application of the cyclic switchover effect in gas sensorics [38, 43] based on the gas sensitivity of Yanson point contacts [49, 50]. Given its nature, the cyclic switchover effect is well-suited for the development of innovative methods for selective detection in liquid media.

The goal of this work is to develop a novel method for detection and differential analysis in liquid media. The proposed method overcomes the limitations of conventional techniques based on classical physics and chemistry principles, offering a new class of sensor devices with atomic-level resolution, high selectivity, and universal applicability for a wide range of sensing applications. This achievement is based on a quantum detection mechanism [38], implemented through changes in the conductance of Yanson point contacts [43] and manifested through the cyclic switchover effect [44].

## 2. Experimental details

The experimental technique and methodology of this work are based on the technology of Yanson point-contact spectroscopy, which allows for easy and rapid execution of all technological operations needed for the creation of quantum point-contact sensors [39, 45]. The research was conducted on a specially assembled stand that included a cell [51] for the preparation of the system of dendritic Yanson point contacts, which serves as the sensing element of the quantum sensor. The dendritic point-contact system was formed on the basis of two electrodes using a modified "needle-anvil" technique developed within the framework of the Yanson point-contact spectroscopy technology [39, 45]. In most experiments, one of the electrodes was represented by an electrochemically sharpened copper needle, while the other was a copper plate (the so-called "anvil"). If necessary, the electrode materials could be changed according to the task being addressed at that moment. In particular, in some experiments, we used electrodes made of zinc. At the beginning of the experiment, the needle electrode was mounted on a spring damper and positioned in the cell perpendicular to the surface of the flat electrode (anvil). The cell was equipped with a device for smoothly bringing together and separating these electrodes with a step of 25 nm [51]. Metallic dendrites were grown under electrochemical conditions at room temperature. Before starting the experiment, the cell was evacuated to a vacuum of $10^{-2}$ Torr. After that, it was filled with argon. All experiments were conducted under atmospheric pressure. A four-probe arrangement was implemented to cancel the influence of current-feeding leads. During the experiment, the tip of the needle was immersed in a droplet of electrolyte, which was the material under study, placed on the anvil. This ensured an electrolytic contact between the electrodes.

By smoothly moving the electrodes, which can be reliably done thanks to the device described in Ref. [51], the needle tip was positioned a few tens of nanometres away from the anvil. In this case, in the current mode, all the electric field applied to the electrodes was concentrated in the interelectrode space. The dendritic Yanson point contact was formed in galvanostatic mode (current held constant) with a current of 20 μA. A detailed description of the methodology for creating dendritic Yanson point contacts can be found in Refs. [39, 43, 44]. The system was powered by a stabilised current source (Keithley 2450). The needle was electrically connected to the negative pole of the source, while the anvil was connected to the positive pole. The moment the current was applied was considered the beginning of the process leading to the emergence of the cyclic switchover effect [44]. The current and voltage were recorded using Keithley DMM6500 multimeters. The data were displayed on a monitor as a dependence of the voltage drop across dendritic Yanson point contact on time, which was automatically converted into a temporal dependence of the contact resistance $R$. The general scheme of the setup is presented in Fig.1. The frequency of voltage recording was 50 points per second, decreasing to 30 points per second when a large volume of information was accumulated. Data recording and processing were carried out using special software developed at B. Verkin Institute for Low Temperature Physics and Engineering of the National Academy of Sciences of Ukraine. The final operation in data processing was the construction of conductance histograms. The methodology for constructing conductance histograms while observing the cyclic switchover effect is described in Refs. [43, 44, 52] and corresponds to the general principles established in the studies on conductance quantisation in Yanson point contacts [53].

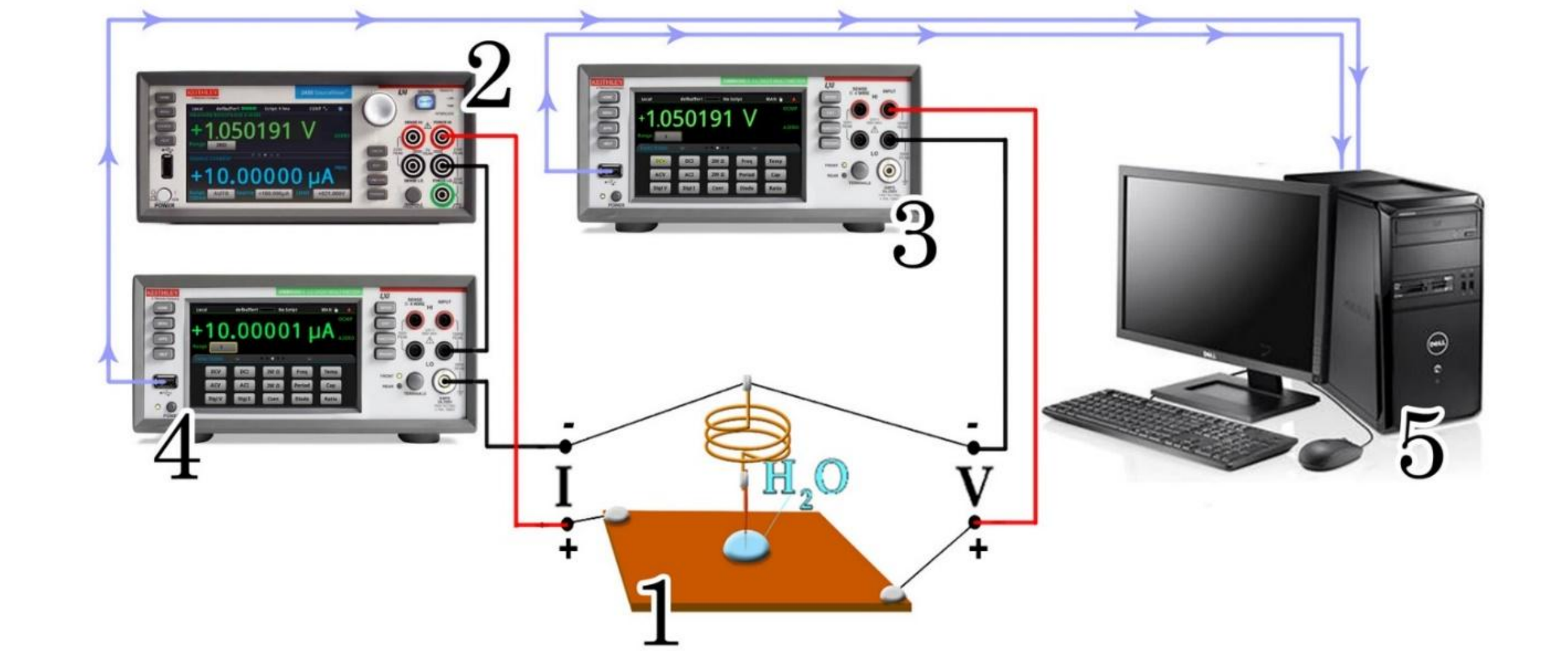

Figure 1. Setup for the dendritic Yanson point contacts study and the cyclic switchover effect observation. 1 – general scheme of the dendritic Yanson point contact preparation; 2 – currents source Keithley 2450; 3, 4 – multimeters Keithley DMM6500; 5 – personal computer.

The study investigated extra pure compounds dissolved in pure water. Both organic and inorganic compounds were used as studied substances. Specifically, measurements were conducted in solutions of copper sulphate, acetic acid, and in liquids containing trace concentrations of lead and zinc. The choice of studied substances was determined by their importance for fundamental research within the proposed approach, as well as their potential practical applications (see section "Results and Discussion").

The condition of the side surface of the needle electrode during the experiment depended on the initial conditions of the research. Since the cyclic switchover effect has an electrochemical nature, the architecture of the electrode system and the state of electrochemically active phase boundaries are absolutely crucial. If at the beginning of the auto-oscillating process, the needle (cathode) was positioned far enough from the anvil (anode), a classical two-electrode electrochemical system emerged. On a scale characteristic of Yanson point contacts [39], this distance is approximately a few microns. In this case, during a prolonged experiment, dendrites formed on the side surface of the needle that did not participate in forming direct interelectrode conduction and creating Yanson point contacts. Dendrites on the side surface of the needle grow much slower than the main dendrites at the needle tip since the electrode polarisation on the side surface, which determines the rate of electrochemical processes, is low. The direction of dendrite growth coincides with the lines of the electric field. However, if at the beginning of the experiment, the tip of the needle was sufficiently close to the counter-electrode surface (on scales characteristic of Yanson point contacts, this is equivalent to several tens of nanometres), such an effect was not observed, and a single dendrite formation occurred (see Fig. 1 in Ref. [51]). Therefore, one of the defining parameters of the initial conditions of the experiment was the interelectrode distance. Minimising this distance was achieved by using the device described in detail in Ref. [51] and the original "know-how" developed by the authors of the present work.

The most thoroughly studied metal for realising the cyclic switchover effect is copper; therefore, detailed research on the detection in liquid media was conducted using copper dendritic Yanson point contacts. During the research, for each analysed concentration of electrolytes and solutions, more than 20 temporal dependencies of the electrical resistance $R(t)$ of Yanson point contacts formed during the cyclic switchover effect were obtained. The resulting curves were used to construct conductance histograms. All groups of histograms obtained for each concentration contained no fewer than 10 histograms with good reproducibility, which were selected through statistical processing of the obtained data. These histograms served as the foundational material for constructing the concentration dependence of the quantised conductance maximum.

## 3. Results and discussion

A crucial prerequisite for observing conductance quantization in dendritic Yanson point contact systems is the cyclic switchover effect [44]. This phenomenon occurs automatically in a two-electrode system (for example, the "needle-anvil" configuration) immersed in an electrolyte through which an electric current is passed. The cyclic switchover effect manifests as a series of structural transformation cycles that repeat sequentially without external intervention. Each cycle consists of a phase of spontaneous dendrite growth, followed by the connection of its tip to the counter-electrode anvil, resulting in the formation of a dendritic Yanson point contact. Subsequently, the conduction channel of the point contact begins to dissolve until complete rupture of the electronic conduction occurs, and the system returns to its initial state, resembling a classical two-electrode electrochemical cell. Once the cycle is completed, all stages of the process are repeated in the next cycle. These cycles can continue for several hours until irreversible changes in the system are observed.

The nature of the cyclic switchover effect is attributed to the emergence of a novel type of nanostructured electrochemical system, which forms on the conduction channel of the Yanson point contact. This system is referred to as a gapless electrode system (GES) [39, 43, 44]. The formation of GES occurs due to a fundamental property of the Yanson point contact, dictated by the specific distribution of potential at the contact during current flow [54]. In the current mode, the voltage drop is concentrated within the conduction channel of the Yanson point contact [39, 49], ensuring the localization of the electric field in a precisely defined nanostructured region. A significant voltage drop across the conduction channel of the point contact leads to an unusual and substantial change in the surface polarization of the channel at the nanometer scale, which is atypical in traditional electrochemistry. The magnitude and sign of polarization are determined by a linear coordinate: the maximum anodic polarization, where oxidation processes occur, is reached at the point of contact between the needle and the anvil surface, while the maximum cathodic polarization, stimulating reduction processes, is observed at the opposite end of the channel. The boundary of polarization inversion, characterized by the absence of directed electrochemical processes, lies between these two ends of the channel. For a homogeneous cylindrical channel, this inversion boundary is positioned at the channel's midpoint [55]. All of these transformations occur over nanometer-scale distances. As a result, a gapless electrode system emerges on the surface of the conduction channel, consisting of electrodes arranged continuously along the main axis of the channel and not separated by an electrolyte layer.

The discovery of GES has spurred the development of an innovative concept in electrochemistry, where a continuous series of electrodes can exist along the metal-electrolyte interface [39]. This distinguishes GES fundamentally from traditional electrochemical systems. In the dendritic Yanson point contact, the GES can be visualized as a continuous "metal-electrolyte" interface along the generatrix of the conduction channel. Due to the specific potential distribution in the Yanson point contact, the GES leads to the formation of anodic and cathodic regions at opposite ends of the conduction channel (see Fig. 1d in Ref. [43]). In the anodic region of the point contact's conduction channel, where the dendrite tip contacts the anvil, oxidative processes are initiated, leading to the dissolution of material from the conduction channel until complete rupture of electronic conduction occurs.

In broad terms, the cyclic switchover effect and the associated transformation processes can be described as follows. During the experimental procedure conducted in current mode within the "needle-electrolyte-anvil" system, dendrite growth is initiated from the needle side. The dendrite tip subsequently contacts the surface of the anvil electrode, resulting in the formation of a dendritic Yanson point contact. Concurrently, a gapless electrode system (GES) forms on the surface of the conduction channel of the point contact, triggering the dissolution of the contact material on the anodic side of the channel and leading to the destruction of the contact. Upon completion of this full set of structural transformations, the system reverts to the state of a classical two-electrode electrochemical cell. As soon as this state is reached, a new cycle begins, and the entire sequence of transformations is repeated.

The structural transformations within the conduction channel of the dendritic Yanson point contact are shaped by the quantum shell effect [39, 44]. The quantum nature of these processes provides the essential conditions for the implementation of a quantum detection mechanism when utilizing point-contact sensors [43]. This phenomenon is particularly evident when studying the temporal dependence of conductance in a dendritic Yanson point contact system during the cyclic switchover effect. To demonstrate this, we consider a typical experimental curve obtained in this study (see Fig. 2). The direct measurement process using a quantum point-contact sensor involves recording the electrical resistance ($R$) of the sensing element as it interacts with the analyte during the cyclic switchover effect (Fig. 2a).

The entire $R(t)$ curve consists of segments, each representing the formation and destruction of a dendritic Yanson point contact in an auto-oscillatory process. This behaviour results from transformations induced by the quantum shell effect during the cyclic switchover effect. The lower resistance levels in the $R(t)$ dependence correspond to contacts with larger diameters, while higher resistance levels indicate smaller diameters of the corresponding contacts. The nanostructured contacts formed during the cyclic switchover effect, characterized by direct electronic conduction, display a resistance range from several ohms to 12.9 kΩ, a value typical of monoatomic contacts [53].

The quantum nature of the transformations in the crystalline structure of the conduction channel of the dendritic Yanson point contact leads to the appearance of bends and steps in the $R(t)$ curve, corresponding to specific conductance values and lifetimes of the respective metastable structures [47]. This can be explained as follows: each metastable crystalline state is associated with a specific conductance, resulting in horizontal sections (steps) in the $R(t)$ dependence (Fig. 2b), which persist for the lifetime of the metastable state [44, 56]. Transitions between crystalline states occur abruptly due to the quantum shell effect. This transition is accompanied by a sharp change in conductance, leading to the emergence of a new step in the conductance curve of the quantum object [57, 58]. The combination of such states, arising during the auto-oscillatory process, forms a conductance histogram [53], which reflects the probability of quantum metastable states within dendritic Yanson point contacts, each characterized by specific values of quantized conductance [43].

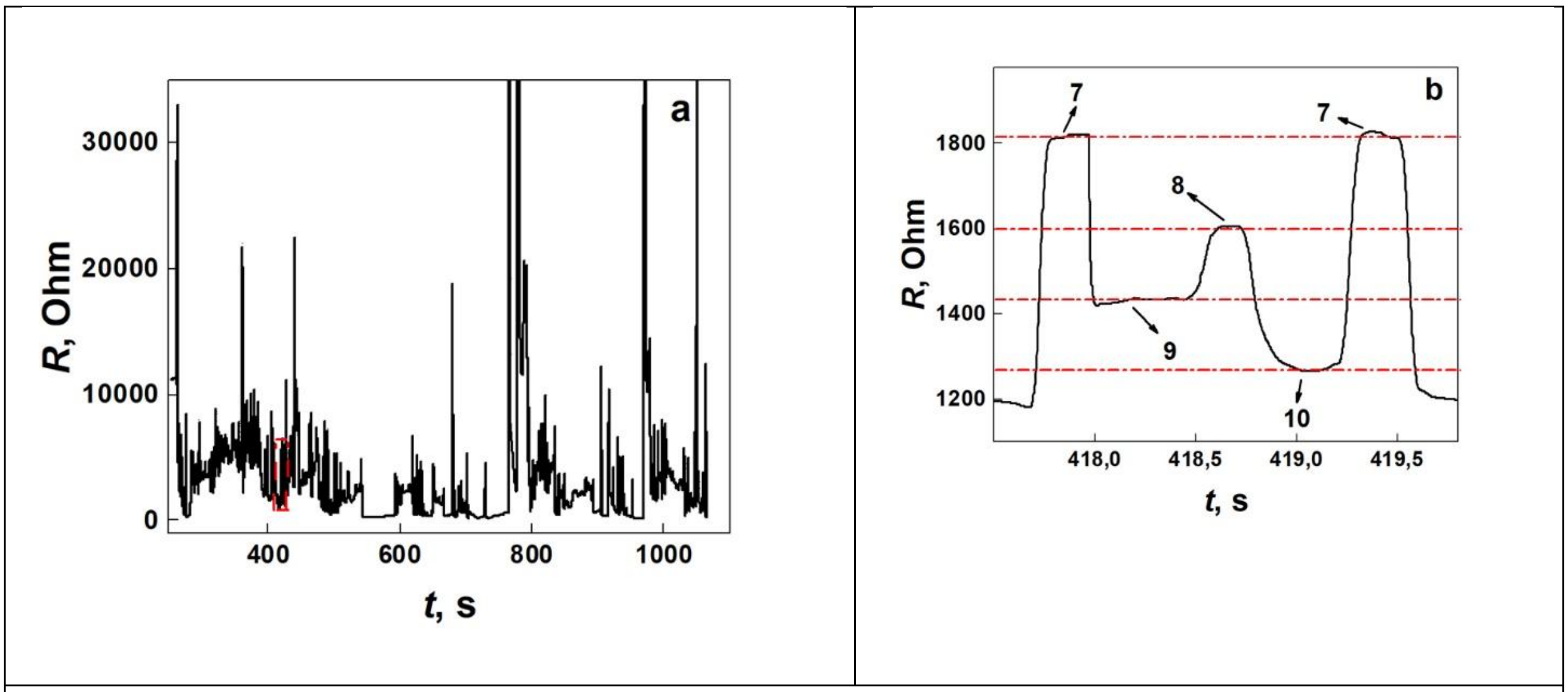

Figure 2. Resistance $R(t)$ dependence of the dendritic Yanson point contact system during the cyclic switchover effect in a solution containing lead ions at a concentration of 50 ppb. (a) Overview of the curve. (b) Enlarged view of the section of the curve indicated by the red rectangle in (a). Arrows show the number of conductance quanta calculated from the resistances of the metastable states. $R$ is the resistance; $t$ is the time.

Depending on the composition of the environment surrounding the dendritic Yanson point contact, the probability distribution for the appearance of metastable quantum states of this nano-object with specific conductance values in the auto-oscillating process undergoes modification [43]. To quantitatively assess the impact of the environmental composition on the configuration of the conductance histogram, it is essential to establish a baseline probability distribution for the metastable quantum states under controlled conditions. As demonstrated in Ref. [43], such a distribution can be obtained experimentally in a standardized environment. Pure water, deaerated with argon to remove dissolved gaseous impurities, was chosen as the standard environment. This choice ensures the purity of the experiment, effectively eliminating the influence of unwanted substances or effects on the recorded results. It is important to note that in bulk form, pure water exhibits ultra-low electrical conductance, which prevents electrochemical processes under normal conditions. However, as shown in Ref. [59], the GES in dendritic Yanson point contacts induces electrochemical reactions in the conduction channel. This is due to the hydration of copper ions on the surface of the conduction channel, which forms a sufficient concentration of ions in the near-surface region, facilitating nanoscale electrochemical transformations.

During measurements, the argon pressure in the cell is maintained at one bar. A conductance histogram characteristic of the inert argon environment has been determined and can serve as a baseline for comparison when studying complex objects. This distribution of probabilities for the occurrence of metastable quantum states in dendritic Yanson point contacts (Fig. 3) is used as reference data to evaluate the interaction of the studied analytes with the quantum sensor's sensing elements. This approach led to the discovery of a novel quantum mechanism for selective detection in complex environments, based on conductance quantization [43], and revealed that each studied system is characterized by its unique "fingerprint" in the form of a distinctive distribution of conductance values for metastable quantum states in dendritic Yanson point contacts.

The histogram obtained in an argon environment is represented by a narrow band of conductance quanta, indicating a deficiency of electroactive components in the studied system.

The ultra-high sensitivity of the quantum sensor enables detection in environments with a resolution of one conductance quantum, i.e., $G/G_0 = 1$ [39], where the conductance quantum $G_0 = 2e^2/h$ is defined through the electron charge $e$ and Planck's constant $h$ [53]. Identifying components in a multicomponent environment using the quantum point-contact sensor can be conducted provided there is an established data library for conductance histograms.

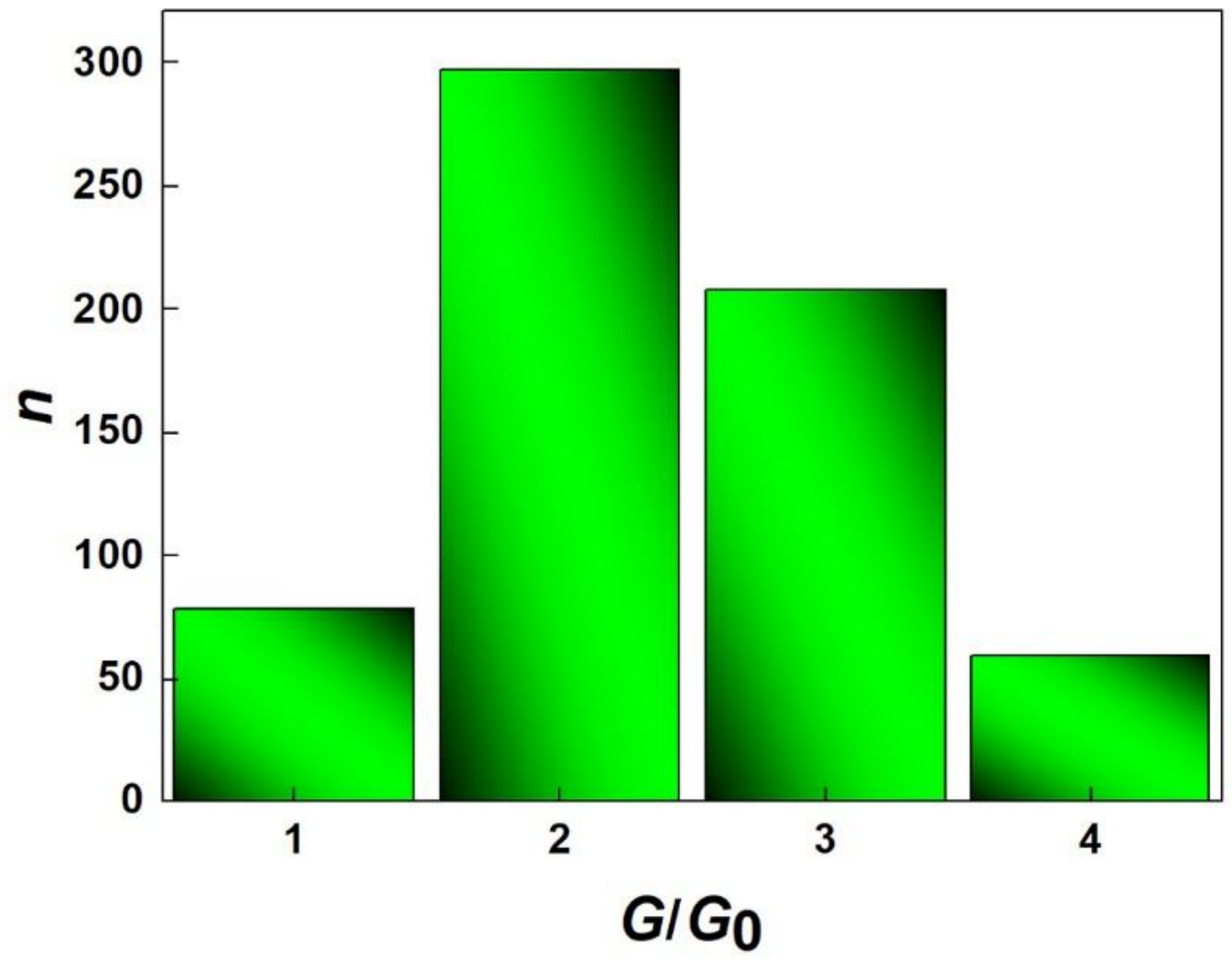

Figure 3. Conductance histogram of dendritic Yanson point contacts in argon-deaerated pure water. $G$ is the conductance, $G_0$ is the conductance quantum, and $n$ is the number of recorded states.

The selective quantum detection method in gaseous environments proposed in Ref. [43] was employed in this study to develop a novel, universal approach for selective detection in liquid environments using quantum point-contact sensors. The new approach was tested with solutions of copper sulphate, acetic acid, and salts of zinc and lead. Copper sulphate was chosen due to its prevalence in galvanostatic copper plating solutions. Additionally, copper is the most widely used metal in Yanson point-contact spectroscopy, allowing us to focus on the critical aspects of the quantum detection mechanism without the need for alternate point-contact nanostructure fabrication techniques. Including metals like lead and zinc further demonstrates the potential of the newly discovered analytical mechanism. The application of this mechanism to develop a reagent-free method for determining trace amounts of heavy metals in water is of significant interest for both fundamental physical science and applied environmental and ecological research. Acetic acid was included to illustrate the method's capability for selectively detecting organic compound impurities in liquid environments.

In studying the influence of ion concentration on the initial characteristics of the quantum sensor, the resistance $R(t)$ of dendritic Yanson point contacts was measured during the cyclic switchover effect across a concentration range of $10^{-1}$ to $10^{-6}$ mol/dm³ for copper sulphate solutions. The resulting curves were used to calculate conductance histograms. Figure 4 presents the comparative conductance histograms derived from these experiments. The high reproducibility of the results, both in the shape of the histograms and in the positioning of their features, is evident in the comparison.

This type of histogram reveals the presence of a specific set of allowed metastable states of the nanostructures, providing evidence of the quantization of conductance in dendritic Yanson point contacts within the studied liquid environment. A particularly noteworthy observation is that the high reproducibility of the conductance histogram for a given electrolyte concentration allows this characteristic to serve as a reliable indicator of the quantum state of dendritic Yanson point contacts for analytical purposes. Confirmed through numerous experiments across various systems, this reproducibility suggests that the conductance histogram uniquely characterizes the studied environment and functions as a generalized criterion for its identification.

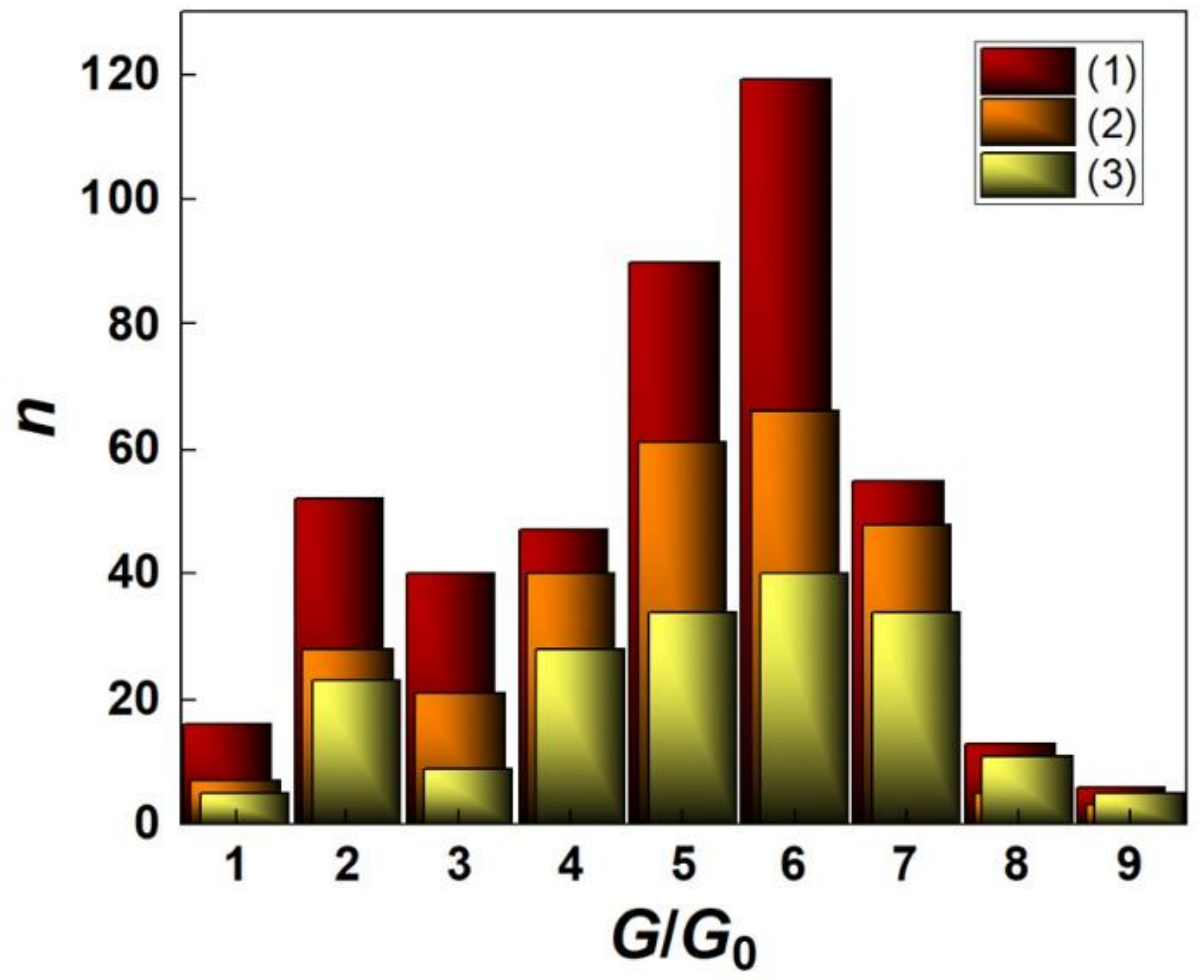

Figure 4. Conductance histograms of copper dendritic point contacts in a copper sulphate solution with a concentration of 0.001 mol/dm³. The calculation was performed for the $R(t)$ dependencies recorded in three experiments conducted under identical conditions. $G$ is the conductance, $G_0$ is the conductance quantum.

Additionally, it is prudent to introduce an additional criterion for selective detection in liquid environments based on the characteristic parameters of the response signal from the quantum point-contact sensor. Histograms corresponding to a specific analyte concentration typically display a prominent maximum (unimodal histograms), corresponding to a state characterized by a consistent number of conductance quanta (Fig. 4). This enables the use of the coordinate of the most intense maximum of the conductance histogram as a calibration parameter for the specific system.

We analysed sets of 10 histograms for each copper sulphate concentration and determined how the position of the histogram's main maximum shifts with varying electrolyte concentration. Using the resulting histograms, we constructed a plot showing the relationship between the quantized conductance of the histogram maxima and the concentration of the copper sulphate solution (Fig. 5). This plot effectively reflects the dynamics of the most probable size of Yanson point contacts synthesized under different experimental conditions.

This dependence can serve as a calibration characteristic for the "copper sulphate – water" system. Using this characteristic, the concentration of an unknown copper sulphate solution can be precisely determined by referring to the abscissa of the maximum on the conductance histogram obtained from the solution. The relationship between the concentration and the histogram maximum can be described by the following equation:

$$C = 4.0 \cdot 10^{-7} \left( \frac{G}{G_0} \right)^{4.2} .$$

To derive this equation, the experimental dependence of the number of conductance quanta at the histogram maximum $G/G_0$ on the solution concentration $C$ was analysed. The results were graphically represented in the coordinates log $C$ (ordinate axis) versus log ($G/G_0$ – 2.60) (abscissa axis). The constant 2.60 was determined by extrapolating the dependence $G/G_0 = f$ (log $C$) to the ordinate axis. In this representation, all experimental data points aligned on a straight line, and the corresponding analytical equation allowed for the extraction of both the exponent and the multiplier in the expression, leading to the equation for determining the salt concentration in the solution.

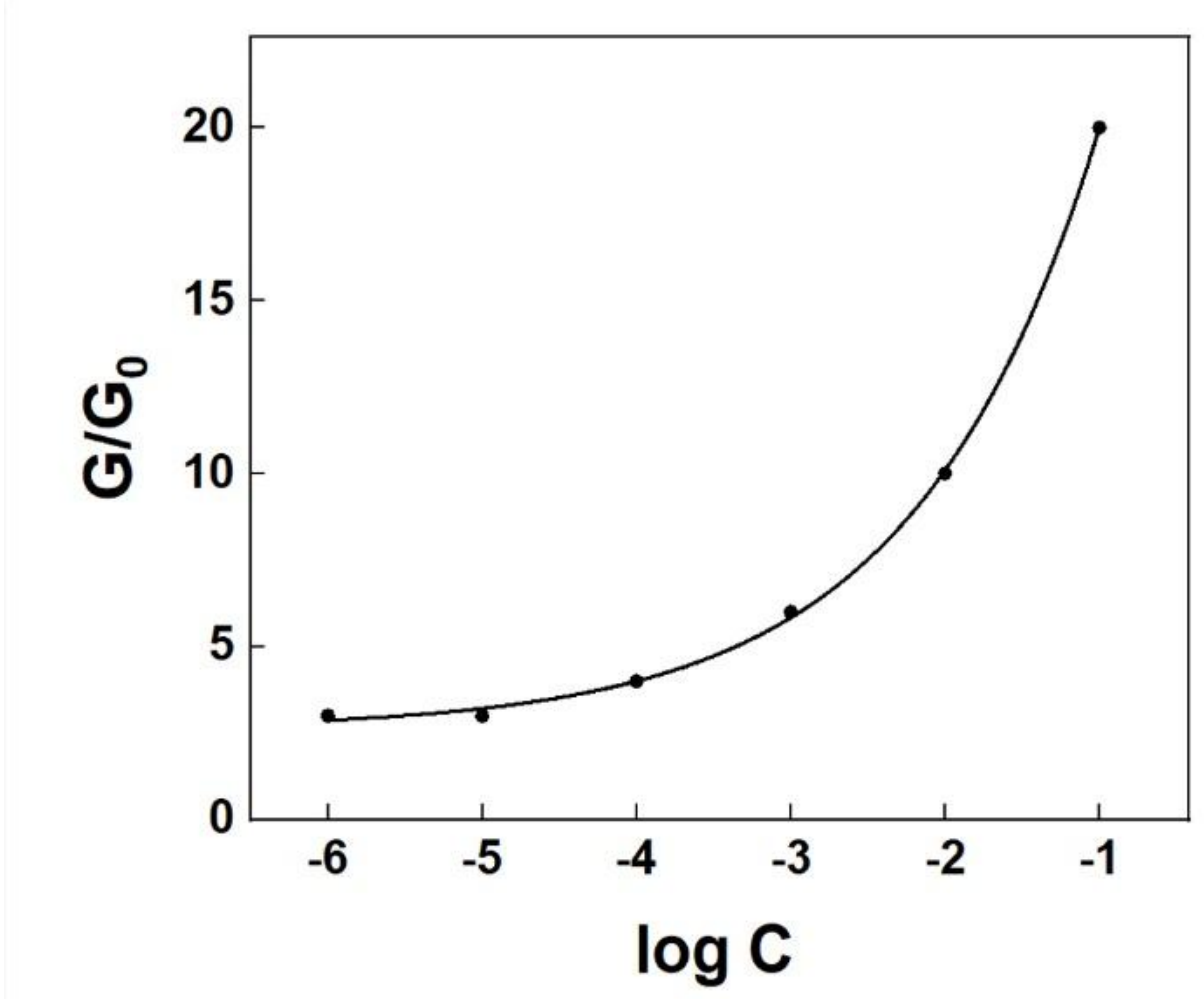

Figure 5. Dependence of the quantised conductance of the histogram maxima on the concentration of aqueous copper sulphate solution. *G* is the conductance, $G_0$ is the conductance quantum, and *C* is the concentration in mol/dm³.

According to the presented data, the higher the concentration of the electrolyte, the larger the diameter of the cross-section of the conduction channel of the dendritic Yanson point contacts formed with maximum probability. The obtained result is in complete agreement with our earlier noted correlation between the sizes of point-contact nanostructures and the concentration of the electrode-active component of the medium in which they were synthesised [47]. Such behaviour of the system can be explained on the basis of the theoretical principles of electrode process kinetics [60]. In a more concentrated solution, the sizes of crystallisation nuclei from which dendrites are formed are larger than in solutions with lower electrolyte concentrations. Moreover, the sizes of point-contact structures and nuclei are comparable, which is why point contacts with larger conduction channel diameters form in saturated electrolytes.

The given example of detection in liquid media corresponds to the case when the ion to be analysed has the same nature as the metal of the conduction channel of Yanson point contact. This case should be considered quite specific, and the list of substances that can be analysed based on the described mechanism is limited to metals capable of forming reversible redox systems. Examples of such electrodes include $Pb/Pb^{2+}$, $Ag/Ag^{+}$, $Tl/Tl^{+}$, $Zn/Zn^{2+}$, $Ni/Ni^{2+}$, $Sn/Sn^{2+}$, $In/In^{+}$, $Cd/Cd^{2+}$. To detect such contaminants, it is sufficient to register the cyclic switchover effect in the corresponding system. In some of the listed metals, quantisation of conductance has already been observed during the creation of dendritic point contacts in an electrolyte through artificial methods [58, 61-63]. Therefore, implementing a cyclic switchover effect and a quantum mechanism for selective detection should not pose any difficulties. In this case, the same approach used for detecting copper with copper dendritic Yanson point contacts can be applied for detecting each of the mentioned substances. The same principle operates here, so the effectiveness of this approach in detecting other heavy metals is expected to be equally high.

To demonstrate this, we conducted measurements of the *R*(*t*) dependence in a system of dendritic Yanson point contacts based on zinc during the cyclic switchover effect. The general principles of the methodology for preparing experiments and observing the cyclic switchover effect in zinc electrode systems correspond to those employed when working with copper electrodes. An aqueous solution of $ZnSO_4$ with a concentration of 0.001 mol/dm³ served as the detectable agent. A typical temporal dependence of the electrical resistance of zinc dendritic Yanson point contacts is presented in Fig. 6.

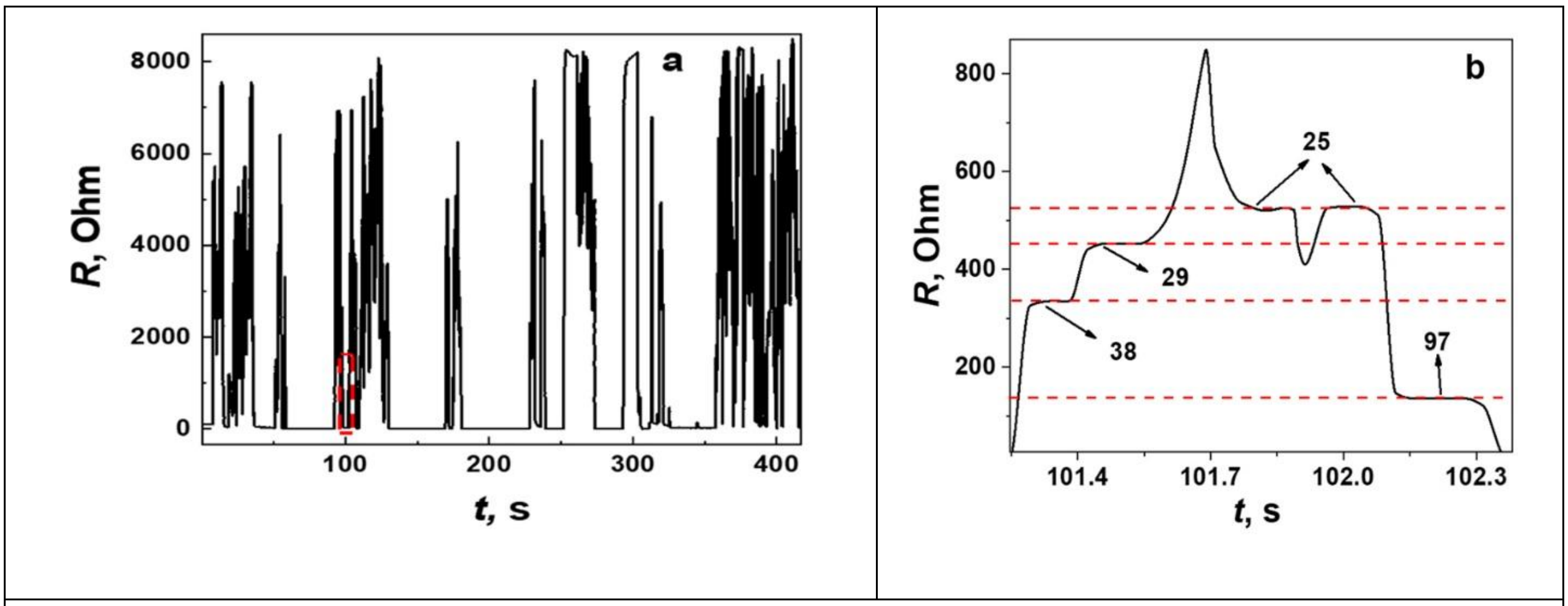

Figure 6. Dependence of the resistance $R(t)$ of the dendritic Yanson point contact system based on zinc during the cyclic switchover effect occurring in a $ZnSO_4$ solution with a concentration of 0.001 mol/dm$^3$. (a) Overall appearance of the curve. (b) Enlarged view of the section of the curve marked by a red rectangle in figure "a". Arrows indicate the number of conductance quanta corresponding to the metastable state. $R$ is the resistance; $t$ is the time.

The overall appearance of this dependence corresponds to the behaviour of the $R(t)$ curves measured in experiments with copper contacts during the cyclic switchover effect. In the case of zinc, a cyclic switchover effect is also present in the current mode: cyclic changes in the conductance of dendritic point contacts are observed, accompanied by quantum structural transformations. The presence of metastable quantum states is well tracked by sections of constant conductance with specific values of conductance, which form steps on the $R(t)$ dependence (Fig. 6b). The value of conductance at sections with specific numbers of conductance quanta can be used to construct conductance histograms. The conductance histogram, as a probability distribution of the occurrence of certain metastable quantum states, serves as a characteristic of the studied system, namely, its energy fingerprint, and allows for the unambiguous determination of zinc ions in a liquid medium using the approach described above for detecting copper sulphate.

Both theoretically and practically, dendritic Yanson point contacts allow for an easy detection of even a single atom of the external agent. This is evidenced by the characteristic appearance of conductance histograms. During the cyclic switchover effect in the ballistic current flow mode, the addition of a single atom leads to a significant change in the conductance of the dendritic Yanson point contact. For example, interaction with a single atom to be detected results in a resistance change as high as at least $12.9/n$ kΩ, where $n$ is the number of atoms in the contact [43, 44]. One conductance quantum corresponds to a change in electrical resistance of the contact by 12.9 kΩ. Such a magnitude can easily be registered using a standard multimeter. Practically, the resolution capability for one quantum is clearly visible in the conductance histogram (see Figs. 3 and 4). Therefore, in the presence of a cyclic switchover effect, the efficiency of detecting single ions of the corresponding metals using zinc contacts or contacts made from other metals will be as high as in the case of copper.

The list of metals capable of forming redox systems, as presented above, is not exhaustive, since the nature of the system can be influenced by the choice of an organic solvent in place of water. To broaden the scope of detectable substances, we performed additional experiments, the results of which suggest that a foreign metal ion can be detected in relation to the material of the conduction channel. This finding positions the quantum point-contact detector as a versatile analytical tool.

In planning and conducting these experiments, it was crucial to eliminate the possibility of contact expulsion of surface copper atoms by the studied ion, as this could rapidly shift the system into a state similar to the base state, where both the conduction channel atoms and the analysed ions are of the same nature. Another important consideration in setting up the experiment was the potential practical application of the results. Bearing this in mind, lead ions were selected as the most suitable analyte to demonstrate the broad capabilities of quantum detection in a liquid medium. Notably, when examining redox reactions in aqueous solutions, lead is more active than copper, thus preventing the so-called cementation process involving the analyte in its oxidized form. Additionally, the particular relevance of analysing lead ions in water arises from the fact that poorly soluble crystalline lead hydroxide, in equilibrium with water, yields lead ion concentrations that exceed the maximum allowable concentration by more than two orders of magnitude (see Ref. [64] and references therein). Therefore, detecting this ion in aquatic environments is critical for environmental monitoring. In light of this, aqueous solutions of lead salts were investigated using copper dendritic Yanson point contacts.

For detecting lead ions, the model environment was prepared using $Pb(CH_3COO)_2$ salt, with a solution concentration of 50 ppb in pure water. It is important to note the conditions necessary for the cyclic switchover effect in environments with low concentrations of the detectable agent. In such cases, there may be a delay in the onset of the cyclic process, with one of the main contributing factors being the state of the metal surface from which the dendritic Yanson point contact is formed in the analysed medium. The concentration of protons in the solution (i.e., the pH level) is a key factor influencing the nature of the metal surface. Alongside acidity, the thermodynamic stability of compounds is determined by the potential of the metal surface. To successfully prepare and conduct the experiment, Pourbaix diagrams, which depict the stability of metal compounds as a function of pH and potential, were used [65]. According to the Pourbaix diagram for copper, maintaining an active metal surface requires keeping the pH of the medium no higher than 4. Additionally, it is essential to control the metal potential, which should not fall below 0.37 V relative to a normal hydrogen reference electrode. Even with the necessary acidity, a decrease in the stationary potential of the copper surface may occur, indicating the formation of surface oxides during the cyclic switchover effect. Consequently, experiments for detecting trace concentrations of lead were conducted in an environment with a pH of -3, maintained by a solution of acetic acid at a concentration of $10^{-3}$ mol/dm$^3$.

In solutions of $Pb(CH_3COO)_2$ salt dissolved in pure water, a cyclic switchover effect was consistently observed. Time-dependent measurements of the resistance of dendritic Yanson point contacts, $R(t)$, were recorded, with a representative example shown in Fig. 2a. These curves exhibit segments with constant conductance, indicative of quantum metastable states in the dendritic Yanson point contacts when immersed in a medium containing trace concentrations of lead ions (Fig. 2b). During the cyclic switchover effect, the metastable states of the conduction channel are periodically reproduced with a certain probability, which can be quantitatively analysed by constructing a conductance histogram.

A typical conductance histogram of dendritic Yanson point contacts in a medium containing lead ions at a concentration of 50 ppb is presented in Fig. 7a. For comparison, a conductance histogram was also generated for dendritic Yanson point contacts immersed in a solution of acetic acid at a concentration of $10^{-3}$ mol/dm$^3$, which contained no lead ions (Fig. 7b).

A comparison of the histograms reveals significant differences in their overall appearance depending on the composition of the analysed liquid medium. In the solution containing lead ions, the peak of the conductance histogram is shifted toward a lower number of conductance quanta relative to the corresponding peak in the acetic acid solution. Moreover, the histogram of the lead ion-containing solution exhibits a distinctive plateau-like feature within the conductance range of 7–14 quanta, a characteristic absent from the acetic acid solution's histogram. The width of the conductance histograms also differs significantly between the two cases.

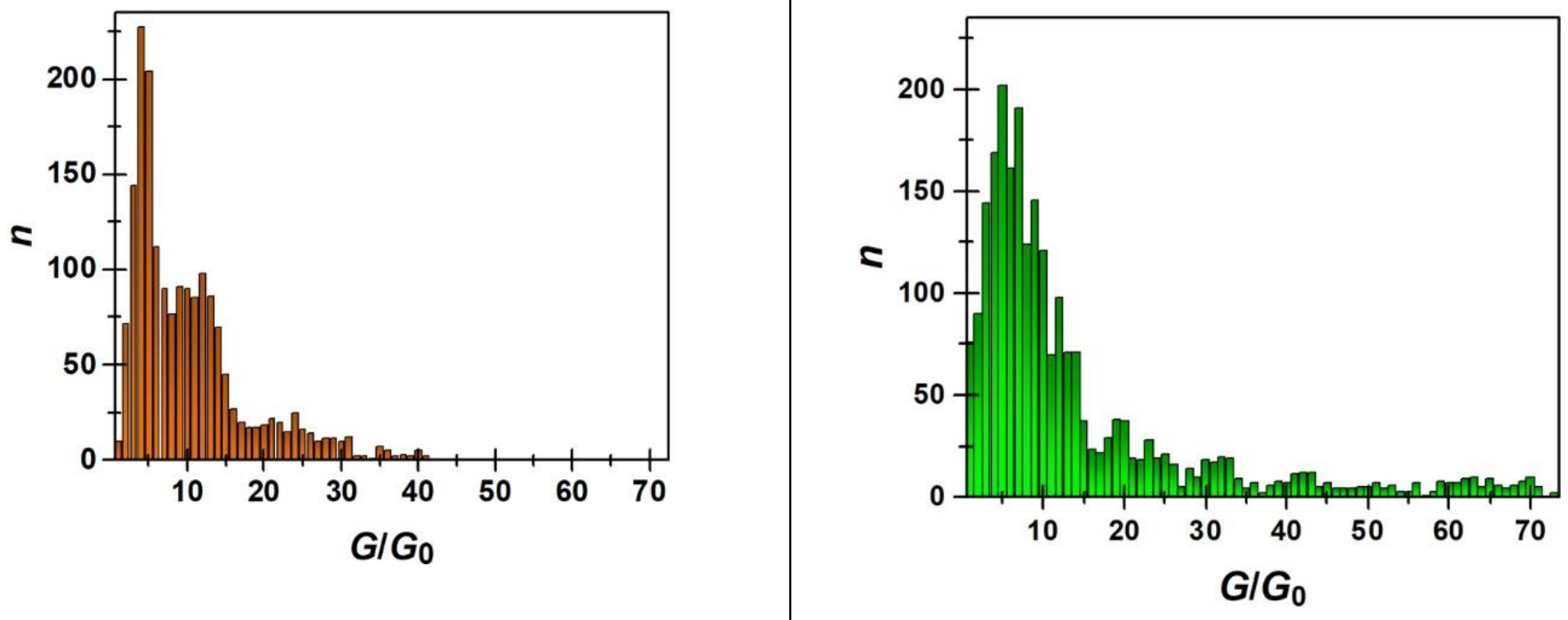

a b

Figure 7. Conductance histograms of copper dendritic Yanson point contacts in liquid media of varying compositions:
(a) Solution of lead acetate with a concentration of 50 ppb.
(b) Solution of acetic acid with a concentration of $10^{-3}$ mol/dm³.

These observations demonstrate that even minor changes in the composition of the analysed liquid medium result in substantial alterations in the conductance histogram of the quantum point-contact sensor's sensing element. This behaviour is consistent with results obtained in the studies of gaseous substances [43], further highlighting the sensitivity of the quantum point-contact sensor to variations in its surrounding medium.

The differences observed in the conductance histograms obtained in the presence and absence of lead ions in a solution containing acetic acid can be explained as follows. In the absence of lead ions, the formation of point contacts is driven by copper ions present in the near-surface layer of the electrode when the tip of a copper needle is immersed in the liquid. A parallel cathodic reaction involves the reduction of protons, resulting in the formation of hydrogen adatoms and molecules. During the reverse polarity phase, the anodic section of the conduction channel may undergo oxidation of hydrogen adatoms.

These reactions occur in a quantum regime, as the dimensions of the system are comparable to atomic scales, and the synthesis of the conduction channel structure is influenced by quantum shell effect. The quantum nature of these processes is reflected in the $R(t)$ dependence as discrete steps, which serve as the basis for constructing conductance histograms.

The introduction of lead ions into the solution gives rise to a competing reaction. During the semi-period of point contact formation, lead ions are reduced alongside copper ions. The reduced lead blocks active centres on the surface, significantly hindering the proton reduction reaction due to the high overpotential of this reaction on metallic lead [60]. As a result, despite the proton concentration in the solution being nearly four orders of magnitude higher than that of lead ions, the proton reduction reaction is largely suppressed. This suppression narrows the range of conductance quanta observed in the histogram, corresponding to the auto-oscillating process in the solution containing lead ions.

Similar phenomena may occur in systems with different compositions or conditions, offering potential for further exploration. Investigating a range of such systems could yield valuable insights with both practical and theoretical significance. This prospect motivates ongoing research, the results of which will be prepared for future publication. These forthcoming studies aim to establish specific calibration dependencies that can be directly applied in analytical practice.

The present work demonstrates the fundamental feasibility of employing a universal quantum point-contact sensor for the selective analysis of liquid media. This method holds promise for active application in detailed analyses of systems relevant to both fundamental and applied research, paving the way for advancements in this emerging field.

The results obtained affirm that the quantum point-contact sensor is capable of detecting trace amounts of lead ions, highlighting the exceptional sensitivity of this novel analytical method based on conductance quantization. Beyond this sensitivity, the presence of lead ions in the solution influences the properties of the double electric layer at the interface between the liquid phase and the conduction channel's surface. For instance, the zero charge potential of a copper electrode in dilute aqueous solutions is approximately +90 mV on the hydrogen potential scale [60]. Experimental studies revealed that the stationary potential of the copper electrode in the investigated solution ranged from 110 to 120 mV, indicating that the unpolarized copper surface is positively charged. Consequently, lead ions are repelled from the Helmholtz plane due to Coulomb interactions.

Under nonequilibrium conditions, where the tip of the copper dendrite is negatively polarized, the polarization can exceed 200 mV. This promotes the accumulation of lead ions in the dense region of the double layer under the influence of Coulomb forces, followed by their reduction. As a result, even minute concentrations of lead ions in the solution can significantly inhibit the proton reduction reaction, as noted earlier.

In conclusion, experiments with both zinc and lead ions underscore the versatility of the proposed method for the selective detection of heavy metals in liquid media via conductance quantization. A key advantage of this approach is its adaptability, enabling tailored solutions for specific analytical tasks. For example, detecting trace concentrations of lead ions using copper dendritic Yanson point contacts in an organic solvent represents a novel application, as such a medium – characterized by a very low content of electroactive agents – was used for sensor studies for the first time. The success of these experiments further demonstrates the universality of the cyclic switchover effect and the robustness of the quantum detection method. This paves the way for developing quantum sensor technologies capable of detecting pollutants in various liquid media across a broad concentration range.

## 4. Conclusion

The results of this study confirm the successful achievement of its objectives by demonstrating the universality of an innovative quantum detection method for identifying diverse agents in liquid media. This method has proven effective across various liquid components and concentrations of target substances. The quantum properties of Yanson point contacts, combined with the cyclic switchover effect as a powerful tool for their activation and measurement, provide a robust foundation for the proposed universal method of selective detection in liquid media.

Moreover, the quantum mechanism of detection, based on conductance quantization, ensures high efficiency and reliability in identifying a wide range of pollutants. The underlying energetic principles of this method provide quantum sensors with absolute selectivity, as supported by previous studies [38, 39]. These attributes make the quantum detection method a highly promising approach for environmental monitoring and other applications requiring precise and selective analysis of liquid media.

The universal method of selective detection in liquid media proposed in this study is implemented using quantum point-contact sensors. These sensors are fabricated on the basis of widely available, conventional metals, eliminating the need for complex and expensive technologies or the synthesis of novel sensor materials. Quantum point-contact sensors, based on dendritic Yanson point contacts undergoing quantum transformations during the cyclic switchover effect, exhibit a range of unique features that set them apart from all previously known analytical techniques relying on changes in electrical conductivity.

One such distinguishing feature is that the sensing elements of the quantum point-contact sensor are automatically and dynamically formed during the measurement process. This is achieved through the special detection mode enabled by the cyclic switchover effect [43, 48]. In this mode, detection involves a sequence of repetitive and simple measurement cycles, each consisting of the creation and dissolution of dendritic Yanson point contacts. In every cycle, a newly formed sensing element with a clean metallic conduction channel surface participates in the measurement. This clean surface facilitates adsorption-desorption processes, forming the sensor's output signal. As a result, time-dependent degradation of the sensing surface is eliminated, ensuring the high reliability of the data obtained.

Unlike traditional conductive sensors that require multiple manual measurement cycles to enhance data reliability and calculate an average value, quantum point-contact sensors automatically perform repeated measurements within the cyclic switchover effect which inherently involves parallel measurement operations [59].

The proposed universal flexibility of this method allows for the optimization of detection strategies based on economic and technical considerations. This adaptability makes it possible to minimize or entirely eliminate reagent use, particularly in applications such as pollutant detection in wastewater. The near absence of reagents in such cases provides significant economic advantages, paving the way for industrial-scale adoption of quantum sensor technologies.

This work demonstrates the successful application of the quantum mechanism of selective detection, based on conductance quantization, in developing a universal method for detecting a wide range of agents in liquid media. The method has proven effective for identifying heavy metals such as copper, zinc, and lead at concentrations as low as a few parts per billion (ppb) and for detecting organic solvents, with acetic acid serving as a representative example. The use of these innovative quantum detection principles opens up vast opportunities for developing a new generation of devices tailored to address critical environmental monitoring challenges.

To enable practical applications of this method in an automated mode, it is essential to establish a comprehensive data library for a broad range of detectable agents. Incorporating machine learning techniques, such as artificial neural networks, into quantum point-contact sensors can further enhance their capabilities. This would facilitate the development of autonomous systems capable of intelligent, real-time operation without human intervention, ushering in a new era of advanced and efficient environmental monitoring technologies.

## Funding

This work was partly supported by the NATO SPS Programme (Ref: SPS.MYP 985481), the MES of Ukraine (Ref: 0123U103762) and the NAS of Ukraine.

## CRediT authorship contribution statement

Conceptualization, G.K., O.P., V.V., N.K., and E.F.; Investigation, A.H., A.S., G.K., O.P., and M.S.; Methodology, O.P., A.S., A.H., and G.K.; Writing – original draft preparation, G.K., O.P., V.V.; Writing – review & editing, G.K., V.V., O.P., M.S., N.K., and E.F.; Visualisation, G.K, O.P., V.V., A.S., and A.H.; Supervision, G.K., E.F.; Project administration, G.K., E.F.; Funding acquisition, G.K., O.P., E.F. All authors provided critical feedback and helped shape the research and manuscript. All authors have read and agreed to the published version of the manuscript.

## Declaration of Competing Interest

The authors declare no conflicts of interest. The funders had no role in the writing or in the decision to publish the manuscript.

## Data Availability

The data are available from the corresponding author upon reasonable request.

*Institutional review board statement*

Not Applicable.

*Informed consent statement*

Not Applicable.